\documentstyle[12pt,epsfig]{article}
\begin{document}
\begin{center}
{\Large \bf Nexus solitons in the center vortex picture of QCD}\\[.5in]
John M. Cornwall*\\
{\em Department of Physics and Astronomy\\
University of California, Los Angeles\\
Los Angeles, Ca 90095}\\[.5in]

{\bf Abstract}\\
\end{center}

\bigskip

\noindent It is very plausible that confinement in QCD comes from linking of Wilson loops to finite-thickness vortices with magnetic fluxes corresponding to the center of the gauge group.  The vortices are solitons of a gauge-invariant QCD action representing the generation of gluon mass.  There are a number of other solitonic states of this action.  We discuss here what we call nexus solitons, in which for gauge group $SU(N)$, up to $N$ vortices meet at a center, or nexus, provided that the total flux of the vortices adds to zero (mod $N$).
There are fundamentally two kinds of nexuses:  Quasi-Abelian, which can be described as composites of Abelian imbedded monopoles, whose Dirac strings are cancelled by the flux condition; and fully non-Abelian, resembling a deformed sphaleron.  Analytic solutions are available for the quasi-Abelian case, and we discuss variational estimates of the action of the fully non-Abelian nexus solitons in $SU(2)$.  The non-Abelian nexuses carry Chern-Simons number (or topological charge in four dimensions).  Their presence does not change the fundamentals of confinement in the center-vortex picture, but they may lead to a modified picture of the QCD vacuum.  \\[.6in]

\footnoterule
\noindent *E-mail address:  Cornwall@physics.ucla.edu\\[.2in]
\noindent UCLA/98/TEP/16 \mbox{} \hfill June 1998
\newpage

\begin{center}
{\bf I. INTRODUCTION}
\end{center}

The center-vortex picture of confinement in QCD was proposed long ago
\cite{c79,th79,mp,no79,y80}.  In essence it identifies the QCD vacuum as a condensate of finite-thickness vortices of co-dimension 2 in d=3,4; these vortices (closed loops in d=3, closed surfaces in d=4) have a finite transverse extension measured by the inverse gluon mass and possess magnetic fluxes which are expressed via elements of the center of the gauge group.  

In more recent times there has been a resurgence of work on the center-vortex picture, primarily on the lattice \cite{t93,kt,fgo,getal,kl}.  This has resulted in spectacular confirmation of the center-vortex picture of confinement, with the $SU(2)$ string tension being fully captured by by the vortex-dominance approximation of replacing the exact Wilson loop value by $(-)^N$, where $N$ is the total linking number of vortices linked to the loop.  

The center vortices resemble Dirac strings in carrying quantized magnetic flux confined to a closed loop or surface, but differ radically from them in that the center-vortex flux is spread out over a finite transverse extent.  If the flux were instead fully localized (to a delta-function in the continuum, or to a single lattice spacing on the lattice) two things would go wrong:  First, the action of the vortices would be quadratically infinite, either in the continuum \cite{c79} or on the lattice \cite{kt}.  Second, these thin vortices would contribute nothing whatsoever to the adjoint ``string tension" (more properly, a perimeter term in the potential resembling a linearly-rising potential which breaks), leaving only a Coulomb term.  In fact, lattice simulations \cite{b82,pt} and continuum theory
\cite{c83,c98,gad} show that the thick vortices of the center-vortex picture do lead to a breakable ``string" potential for the adjoint representation, directly attributable to the thickness of the vortices which can only partially overlap the Wilson loop.

Avoidance of Dirac-string singularities is a key point in constructing solitons in the continuum center-vortex picture, and will be important for us in constructing the objects considered here, which we call nexus solitons.  A nexus soliton has several thick vortices (up to $N$ of them for gauge group $SU(N)$) which meet at a central region which is essentially of magnetic-monopole character, with exponentially-falling field strengths as one moves out from the center.  In order that this central region have finite action, the magnetic fluxes of the several vortices meeting at the center should add to zero.  If not, the monopole region has fields which diverge like $r^{-2}$ at the origin, like the Wu-Yang monopole or the broken-symmetry monopoles found by Corrigan {\it et al.} long ago \cite{cofn}.   Moreover, if the fluxes do not add to zero there can be {\em long-range} monopole fields behaving like $r^{-2}$ at large distances, as well as naked Dirac strings.  These last two phenomena are due to the termination of Dirac strings at the monopole, in the case where the fluxes do not add to zero (mod $N$).

In previous solitons of the center-vortex picture \cite{c79} this unwanted behavior is simply avoided, in part by having the loop or surface representing the vortex be closed, and in part by cancellation of Dirac-string singularities between a transversely-extended part of the solution and a naked Dirac string.
(In fact, the center vortex is simply an Abelian Nielsen-Olesen vortex embedded in the gauge group but with infinite Higgs mass, or otherwise described it is a soliton of an effective low-energy QCD action consisting of the usual $G^2$ term plus a gauged non-linear sigma model term, giving a gauge-invariant mass to the gluon.)
Any boundaries of the loop or surface gives rise, as is well-known, to objects mathematically identical to the original Dirac monopole plus associated string.
In addition to these closed vortices, the center-vortex picture also has a spherically-symmetric static solution, resembling a sphaleron but with no symmetry breaking \cite{c76}.  It has no Dirac strings in the usual spherical gauge presentation.\footnote{Like the Wu-Yang monopole it can be transformed into an Abelian gauge where it does have a Dirac string, but the required gauge transformation is singular.}

The existence of nexus solitons was alluded to long ago \cite{c79,ct}, but to the author's knowledge these have not been explicitly constructed.  We partially fill that gap here, both for $SU(2)$ and $SU(3)$.  In certain cases, which we call quasi-Abelian, we can give explicit analytic results and in other cases, which are fully non-Abelian, we restrict ourselves to variational estimates.  These nexus solitons go considerably beyond the original vortex solutions, which as mentioned above are really imbedded Abelian Nielsen-Olesen vortices; the new solitons really cannot be described by Abelian embeddings.  The non-Abelian solution which we discuss is essentially a QCD sphaleron \cite{c76} from which several fat vortices emerge. It carries topological charge (or Chern-Simons (CS)  number in d=3).  It can be thought of as a sphaleron-vortex composite, and in $SU(2)$ is energetically stable against decay into a sphaleron and a vortex.  
In $SU(3)$ the quasi-Abelian nexus with three unit-flux vortices is stable against decay into a QCD sphaleron on energetic grounds; this stability is not likely for the $SU(3)$ non-Abelian nexus, which is much heavier.   

It is worth noting \cite{c94} that even the original Abelian center vortices can carry topological charge, which in d=3 is associated with the linking or self-linking of vortices (in d=4, this is expressed as an intersection number of surfaces).  This topological charge is not necessarily integral, but can be quantized in units of $1/N$ for gauge group $SU(N)$.  This need not be a contradiction with periodicity of $2\pi$ in the $\theta$ angle, or with integral quantization of the Chern-Simons level.  

Aside from questions of $\theta$-dependence of QCD dynamics, which we will not discuss here, of what good are these vortices?  The answer is, for confinement at least, that they are not good for much of anything; they do not \cite{c79} change the basic picture of confinement for the fundamental representation which arises when one uses center vortices without nexus solitons.  An interesting question which we are not in a position to answer yet is whether nexus solitons, which are capable in principle of internetting the center-vortex vacuum in a very complicated way (see Fig. 4(b) below), actually do so.  Put another way, the question is whether the change in entropy arising from this internetting is in a sense and of a size which can over come the positive action needed for the nexus monopole regions.  If this happens the vacuum condensate will look quite different from the usual (naive) center-vortex condensate, and possibly could be detected in lattice calculations.  We will discuss in the concluding section what could be learned about nexus solitons and other elements of the center-vortex picture from the lattice.    

We begin with a quick review of the elements of the center-vortex picture, including the effective action.

\bigskip

\begin{center}
{\bf II.  THE CENTER-VORTEX PICTURE}
\end{center}
For convenience we will from now on, except when otherwise noted, work in three dimensions, since all the solitons we find are static.

It was argued long ago \cite{c76} that the low-energy effective energy functional $E$ of QCD consisted of the usual three-dimensional action plus a gauged non-linear sigma model term, representing the generation of a gluon mass.  In turn, the gluon mass appears \cite{c82} because of the fundamental infrared instability of QCD in d=3,4, where the Schwinger-Dyson equations only have non-singular solutions when a mass of sufficient size is generated.  The energy is:\footnote{We use the usual anti-hermitean gauge potential matrix $(\lambda^a/2i)A^a(x)$, with $Tr\lambda^a\lambda^b=(1/2)\delta_{ab}; \;\;D_i=\partial_i+A_i$.  The matrix $U$ is a unitary matrix undergoing left transformations when the gauge potential is gauge-transformed.}  
\begin{equation}   %1%
E=\frac{1}{g^2}\int d^3x \{ \frac{-1}{2}Tr [G_{ij}(x)]^2 +M^2Tr |D_iU|^2 \}.  
\end{equation}
This action is usable only at low energies, because at high energies we must account for the vanishing \cite{c82,la} of the mass, something like 
$M^2(q)\sim \langle G^2 \rangle /q^2$ where $\langle G^2 \rangle$ is the condensate expectation value.\footnote{We will not actually account for this high-energy decrease in mass, with the consequence that the second term in the action (1) is logarithmically divergent at short distances, for the vortices.  Of course, if the mass vanishes quadratically at short distances in actuality, this is a spurious divergence.  In fact, the nexuses themselves do not suffer from this logarithmic divergence.}  The mass must vanish at large momentum if we are to find solutions to the Schwinger-Dyson equations \cite{c82}.  

The fundamental Abelian-vortex solution of the equations of motion coming from
the equations of motion of (1) is:  
\begin{equation}    %2%
A_i(x;J)=(2\pi Q_J/i)\epsilon_{ijk}\partial_j\oint        dz_k\{\Delta_M(x-z)-\Delta_0(x-z)\}                                           \end{equation}
Here the integral is over a closed loop of coordinates $z$,$\Delta_M\;(\Delta_0)$is the massive (massless) free propagator in three dimensions, and $Q_J$ is an $N\times N\;\;SU(N)$ matrix in the Cartan subalgebra.  It is normalized so that the long-range ($\Delta_0$) part of the Wilson loop integral $P\exp \oint dx_kA_k(x)$ for a Wilson loop linked once to the vortex is in the center of the group.  For the fundamental-representation Wilson loop, this means
\begin{equation}                                       %3%
\exp(2\pi iQ_J)=\exp (2\pi iJ/N).
\end{equation}
A convenient set of matrices $Q_J$ is formed from a set of fundamental Cartan matrices $Q_i$:
\begin{equation}                                     %4%
Q_i= diag(1/N,\dots 1/N, -1+1/N, 1/N,\dots 1/N)
\end{equation}
where $-1+1/N$ is in the $i$th position. Only $N-1$ of these matrices are linearly independent, because the sum of all of them is zero.  Then the matrices $Q_i+Q_j+Q_k+\cdots$, with $i\neq j\neq k\neq \cdots$, obey (3) if there are
$J$ terms in the sum.  For $1\leq J\leq [N/2]$, where $[\cdot ]$ indicates the integral part, we can choose in any convenient way one representative of $Q_J$ from such sums, and for greater values of $N$ we use instead the matrices
$Q_{N-J}\equiv Q_{-J}$, representing anti-vortices of flux $-J$.  For future use we record:
\begin{equation}    %5%
TrQ_J^2=\frac{J(N-J)}{N}
\end{equation}

Confinement now arises in d=3,4 \cite{c79} by averaging over the fluctuating phase factors, equation (3), which arise in evaluating the expectation value of the Wilson loop.

For $SU(2)$ there is only one non-trivial vortex, and it is self-conjugate; for $SU(3)$ there is one vortex and its anti-vortex.

\bigskip

\begin{center}
{\bf III. NEXUS SOLITONS}
\end{center}

We are looking for solutions to the equations of motion coming from (1) which have several thick vortices meeting at a monopole-like center, or nexus, with finite action and non-singular field strengths.  For $SU(2)$ the only non-trivial case is when two $J=1$ vortices meet, as in Fig. 1.  By convention we choose ingoing arrows, representing vortex magnetic fields directed toward\footnote{Defining the sense of magnetic fields in non-Abelian theories requires some conventions.  We choose to assign a positive sign to all the $Q_J$ and a negative sign for $Q_{N-J} \equiv Q_J$ for $1\leq J \leq [N/2]$.} the nexus, to describe a nexus, and the opposite direction corresponds to an anti-nexus.  For $SU(3)$ there is only one additional non-trivial nexus, as given in Fig. 2, in which three $J=1$ vortices meet.

Actually, these figures are slightly misleading, since ultimately the vortex strings must close.  The true situation, therefore, is that of Figs. 3 (for $SU(2)$) and 4 (for $SU(3)$) and generalizations thereof, which have as many nexuses as anti-nexuses.  We note a device for constructing pictures of nexuses:  the $SU(3)$ graphs of Figs. 2 and 4 are topologically the same as baryonic Wilson loops, with baryon violation of $\pm$1 unit at every nexus or anti-nexus (given that the baryon number of a quark is 1/3).  Of course, such B-violation is entirely absent in QCD, which is vector-like, and this is only a heuristic analog.

We begin by looking for nexuses which are essentially Abelian, like the center vortices themselves.    There are two possible types:  $U(1)$ nexuses, and quasi-Abelian $SU(N)$ nexuses with charges in the Cartan subalgebra.  In both cases the only possible nexuses are constructed of thick vortices whose fluxes add to zero at the nexus core.  However, the $U(1)$ nexuses are marginally stable to decay into simple loops, which is not so for the quasi-Abelian nexuses.

It will be evident that the essentially Abelian nexuses we find in this section are in some sense composites of monopole-like excitations.  However, it will turn out to be impossible to separate the nexuses into their fundamental parts, since each part is a monopole with a thick vortex carrying the flux into the central monopole region, but a naked Dirac string is needed to carry the flux out.  In a nexus all these naked Dirac strings cancel because of the condition that the total flux is zero; they do not cancel for the pieces from which the nexus is made.
  
\bigskip

\begin{center}
{\bf A.  U(1) nexuses?}
\end{center}
We ask whether there is a nexus
in a $U(1)$ gauge group, like electromagnetism, but with a mass term; the physical analog would be a superconductor.  It should be apparent that all the vortex fluxes must add to zero at the nexus, or else there will be monopole fields which are either singular at short distances or are long-range or both. However, for future use it is worth detailing the explicit forms of the difficulties encountered with non-zero total flux.  We first give the well-known  form of the solution (2) for an Abelian  center vortex lying along the $z$-axis, in standard cylindrical-coordinate notation: 
\begin{equation}     %6%
A_i(x)=\frac{Q_J}{i}\hat{\phi}_i[MK_1(M\rho)-\frac{1}{\rho}].
\end{equation}
Here $K_1(M\rho )$ is the Hankel function of imaginary argument, which falls off exponentially at large argument but behaves like $1/M\rho$ near $\rho=0$.  This term comes from the $\Delta_M$ propagator in (2), while the $1/\rho $ term comes from $\Delta_0$; this latter term is a long-range pure-gauge term, obviously singular at $\rho =0$.  The $K_1$ term just cancels the explicit $1/\rho$ term.  This is what we meant earlier when we said that the center vortex has an extended part with a Dirac string just cancelling an explicit Dirac string.  If this did not happen there would be a field strength proportional to $\delta (x)\delta (y)$, which yields a quadratically-divergent action.  Clearly this cancellation occurs for any closed-loop integral in (2), since $\Delta_M$ behaves like $\Delta_0$ at short distances.  

Now consider terminating the string integral in (2) at some convenient point, say the origin.  Then the $\Delta_M$ part of (2) gives a thick vortex along the negative $z$-axis plus a massive monopole whose (isotropic) field strength goes like
$1/r^2$ at the origin but decays exponentially at infinity.  The $\Delta_0$ part gives the original Dirac monopole, with its string cancelling that of the massive term as before, but also with a {\em long-range} monopole with field strength $\sim 1/r^2$ at both short and long distances.  The short-distance part of this field strength is cancelled by that of the massive term, but the long-distance part survives.  This does not happen in QCD, where all field strengths (but not all gauge potentials!) are short-ranged, so it would seem that the $\Delta_0$ part of (2) must have an integral over a closed loop.  But then if we terminate the massive part of (2) at some point, its short-distance field behavior of $1/r^2$ is uncompensated, and leads to a badly-divergent action.  Moreover, there is now a naked Dirac string coming from the long-range part of ((2).  All of this is unacceptable.

Of course, the way to cure these problems is to make sure that the sum of vortex fluxes is zero, whenever several strings terminate at at common point.  The gauge potential is of the form
\begin{equation}     %7%
A_i(x)=\frac{2\pi}{i}\epsilon_{ijk}\partial_j\sum_A [q_A\int_{\Gamma(A)}dz_k(\Delta_M(x-z)-\Delta_0(x-z))]
\end{equation}
where the sum of the $U(1)$ fluxes $q_A$ is zero, and the $\Gamma_A$ are string paths which meet at the origin, as in Fig. 5.  Each path integral runs from 0 to infinity.  All the {\em long-range} monopole fields from the $\Delta_0$ terms cancel, because they are isotropic and string-independent; the short-range parts of these terms cancels the $1/r^2$ field strengths from the massive terms at the origin, for the same reason; and the Dirac strings all cancel between the short-range and long-range terms, just as for the center vortex.

Note that this is not useful as a model for $SU(2)$ nexuses, since with two fluxes $q_1=-q_2$ the fields do not correspond to Fig. 1, but rather one field points into and the other out of the nexus.  This just gives the usual center vortex with no nexus, at least if the strings joining at the nexus lie along a common axis, say the $z$-axis.  The $SU(2)$ nexus is truly non-Abelian, and will be discussed in Section IV.    

There is another more useful interpretation of the gauge potential (7), in which the strings form closed loops, some segments of which lie along a common line.
Examples of triple nexuses, corresponding to $SU(3)$, are given in Fig. 6, which may be considered as possible realizations of Fig. 4.  In Fig. 6(a), if the strings labeled 1,2 have incoming fluxes $q_1,q_2$ respectively, which we will take to be positive, then the double string has incoming flux $-q_1-q_2$ and automatically respects the condition of zero flux sum.  However, the configuration of Fig. 6(a) is manifestly  unstable into decay into its consituent loops, because the action of the double string, proportional to $(q_1+q_2)^2$, is larger than the sum of the actions for strings 1 and 2.

In Fig. 6(b), if the string fluxes $q_i$ are chosen to add to zero, then the triple string down the middle is a phantom string, with no energy, because its flux is zero.  This configuration is not unstable to decay into its constituent loops, which would require supplying the energy needed to separate the phantom loop into three real loops.  However, it is not a useful description of true $SU(3)$ nexuses, because these will have the same action (per unit length) along each of its three vortices.  In a $U(1)$ group where the $q_i$ are three real numbers, it is impossible to find three real non-zero numbers whose sum is zero
and whose squares are all equal.

We conclude that a $U(1)$ description of nexuses is not possible, and turn to quasi-Abelian nexuses.

\bigskip

\begin{center}
{\bf B.  Quasi-Abelian nexuses}
\end{center}

The only interesting $SU(2)$ case is the non-Abelian case of the next section, so we start with $SU(3)$.  We use the form (7)    
\begin{equation}     %8%
A_i(x)=\frac{2\pi}{i}\sum_{a=1}^3 [T_{aa}\int_{\Gamma(a)}dz_k(\Delta_M(x-z)-\Delta_0(x-z))]
\end{equation}
where $T_{11},T_{22},T_{33}$ are defined in the Appendix; these matrices, whose sum is zero, are just the flux matrices $Q_i$ of equation (4) for $SU(3)$ but with different names.  They obey
\begin{equation}                                       %9%
\exp(2\pi iT_{aa})=\exp (2\pi i/3)\;\;(no\; sum\; on\; a).
\end{equation}
The strings $\Gamma_a$ in (8) correspond to the numbered strings in Fig. 6(b).
As in Section IIIA the triple string in this figure is a phantom string, with no Dirac string or thick vortex, so it contributes nothing to the action.  

We will compute the action of the nexus, that is, the total action associated with (8) minus the action associated with the vortices.  Only the special case where the non-trivial part of string $a$ runs along the positive coordinate axis $a$ all the way to infinity will be given explicitly.  In evaluating the action from (1) observe that the $Tr|D_iU|^2$ term is the same as $Tr|A_i-U\partial_iU^{-1}|^2\equiv Tr|{\mathcal A}_i|^2$, and that this form involves only the short-range $\Delta_M$ part of $A_i$; the $\Delta_0$ part comes from $U\partial_iU^{-1}$.  This mass part of the action contains the short-distance logarithmic divergence mentioned earlier, but it occurs only in the thick vortices and not in the nexus, which has finite action.

Taking the traces in the action (1) yields the action 
\begin{equation}   %10%
I=I_{(12)}+I_{(13)}+I_{(23)}
\end{equation}
where
\begin{equation}     %11%
I_{(ab)}=\frac{1}{3g^2}\int d^3x\{ [\vec{B}^{(a)}-\vec{B}^{(b)}]^2
+M^2[\vec{{\mathcal A}}^{(a)}-\vec{{\mathcal A}}^{(b)}]^2\}
\end{equation}
and
\begin{equation}     %12%
B_i^{(a)}=M^2\int \frac{d^3q}{(2\pi )^2}\frac{e^{i\vec{q}\cdot \vec{x}}}
{(q^{(a)}-i\epsilon)(q^2+M^2) }(\delta^{i(a)}-\frac{q^iq^{(a)}}{q^2}),
\end{equation}
\begin{equation}     %13%
{\mathcal A}^{(a)}_i=\int \frac{d^3q}{(2\pi )^2}\frac{e^{i\vec{q}\cdot \vec{x}}
\epsilon_{ij(a)}q^j}{(q^{(a)}-i\epsilon)(q^2+M^2)}.
\end{equation}
In these equations, $q^{(a)}$ and similar forms means the appropriate component along the axis $a$ of the chosen string; {\it e.g.}, $q^{(a)}=\vec{q}\cdot \hat{e}^{(a)}$, where $\hat{e}^a$ is the unit vector along the string direction (assumed to be straight).  In our case, with the string along the axes, one can drop the parentheses and interpret the $a$ as ordinary spatial indices.  The $i\epsilon$ serves as a cutoff for the integration out to infinite distance along an axis, that is, one can think of the integration along an axis being stopped at a distance $L=\epsilon^{-1}$.  The total action in (10) has a part diverging like $\epsilon^{-1}$; we will define the nexus action as this total action minus the divergent part.

Straightforward calculation gives the total energy of the nexus as
\begin{equation}    %14%
I=\frac{2\pi M}{g^2}.
\end{equation}
This is a remarkably small value, about one-tenth of the energy \cite{c76} of the QCD sphaleron.  But unlike the sphaleron this quasi-Abelian nexus carries no topological charge, as is easily checked.  We will see, when we consider fully non-Abelian configurations next, that forming topological charge costs a considerable amount of energy.  The energy of a nexus depends on the orientation of the strings, and  
the value in (14) is neither the smallest nor the largest energy as the orientations vary, but it is characteristic.  Clearly, it is possible for the whole collection of vortices plux nexus to annihilate itself, by choosing the strings to coincide everywhere, but it is also possible to go to zero nexus energy in another way.  Imagine taking strings 1 and 2 of Fig. 6b to lie along the negative $z$-axis, while string 3 lies along the positive $z$-axis as we started with.  The configuration is geometrically like that of Figs. 1 or 3 now.  However, since $T_{11}+T_{22}=-T_{33}$, one of the strings is reversed and this configuration turns out to be an ordinary center vortex with no nexus at all.

\begin{center}
{\bf IV.  NON-ABELIAN NEXUSES}
\end{center}

The only interesting nexus in $SU(2)$ gauge theory is not a quasi-Abelian one, which is just like the $SU(3)$ nexus described above with strings 1 and 2
coinciding.  It is a truly non-Abelian object, roughly describable as a deformation of the $SU(2)$ sphaleron mentioned above \cite{c76}.  There are also $SU(3)$ non-Abelian counterparts.  In both cases there is no analytic solution, and it would be tedious to find good numerical solutions, since one must solve several non-linear partial differential equations simultaneously (compare to the $SU(2)$ sphaleron, which is described by a single non-linear ordinary differential equation).  So we will content ourselves with some simple variational estimates for the nexus energy in $SU(2)$; for $SU(3)$ (and by a simple extension $SU(N)$) we stop at setting up the (non-trivial) gauge kinematics for the nexus.  The primary reason, aside from the complexity of
going further, is that it is clear that the non-Abelian nexuses have considerably higher energy than the quasi-Abelian nexuses and are not likely to play an important role.   

\begin{center}
{\bf A.  SU(2) non-Abelian nexus}
\end{center}

Although the nexus has cylindrical symmetry only, it is useful to think of this nexus in relation to the usual Witten spherically-symmetric static ansatz
\begin{equation}   %15%
iA_j=\frac{1}{2}[(\Phi -1)X^*_i +c.c.]+\hat{r}_jJ\cdot \hat{r}B_1;
\end{equation} 
\begin{equation}    %16%
X_j=\epsilon_{jak}J_a\hat{r}_k+i(J_j-\hat{r}_jJ\cdot \hat{r}).
\end{equation} 
Here $J_k=\sigma_k/2$ are the $SU(2)$ generators and $\Phi,B_1$ are functions of $r$.  As usual $\Phi\equiv \Phi_1+i\Phi_2$ is a complex scalar field coupled to a $U(1)$ gauge potential $B_{\gamma}$ whose $r$-component is $B_1$.

We can motivate further developments by referring to some well-known results on the Wu-Yang monopole, which is recovered at $\Phi,B_1=0$ in (15).  This monopole, because of a singularity at the origin, is not physically interesting, but it is mathematically simple.  As is well-known this monopole can be transformed, by a {\em singular} gauge transformation, to an Abelian form with a Dirac string:
\begin{equation}   %17%
A^{\prime}_j=UA_jU^{-1}+U\partial_jU^{-1}=\frac{i}{\rho}J_3\hat{\phi}_j(1-\cos \theta )
\end{equation}
\begin{equation}    %18%
U=\exp i\theta J\cdot \hat{\phi};\;UJ\cdot \hat{r} U^{-1}=J_3.
\end{equation}
This form is essentially the original Dirac monopole, with a string along the negative $z$-axis.  The string flux is twice the minimum flux (corresponding to $J=2$ in the notation of equation (3)), so it contributes nothing to Wilson loops.  

By a second singular gauge transformation with $V=\exp i\phi J_3$ we can shift half this flux to the positive $z$-axis (which replaces $1-\cos \theta$ by
$-\cos \theta$ in (17).  At this point the magnetic fields associated with the vortex cores do point as desired (see Fig. 1). We can now, if we wish, make another gauge transformation with $U^{-1}$ to restore spherical symmetry as much as possible.

The combination of these three gauge transformations, namely
\begin{equation}   %19%
U^{-1}e^{i\phi J_3}U=e^{i\phi J\cdot \hat{r}}\equiv W,
\end{equation}
when applied to the Wu-Yang configuration, leads to a deformed kinematical ansatz which we give below in equation (21).  Within this new ansatz we can describe the conventional QCD sphaleron \cite{c76,ct} and deformations which describe thick vortices emerging from it.  

The first step in implementing the new ansatz is to determine the appropriate pure-gauge behavior of the nexus-vortex combination at large distances (by which we mean that both $\rho$ and $z$ are large).  We compose this from the gauge $W$ in (19) above, appropriate to represent the strings pointing into the nexus of Fig. 1, and a pure-gauge form appropriate to a spherical configuration, such as a sphaleron.  So we take the behavior at long distance of the gauge potential to be:
\begin{equation}  %20%
A_j \rightarrow Z\partial_jZ^{-1},\;Z=e^{i(\beta(r)+\phi )J\cdot \hat{r}}.
\end{equation}

We now wish to apply these considerations to deform the usual QCD sphaleron
\cite{c76,ct}.  This object is spherically-symmetric, has short-ranged field strengths, and a CS number of 1/2.  It corresponds to choosing $\beta (r)$ in equation (20) to be $\pi$.  The deformation we use is to be thought of as an ansatz for determining approximately the nexus mass by a variational principle; the ansatz is not necessarily one which fully expresses the kinematics of the nexus.\footnote{It is easy to write a cylindrically-symmetric ansatz which is kinematically self-consistent.  This leads to coupled non-linear partial differential equations for nine scalar components, which we have not studied in detail.}  So we write  
\begin{equation}   %21%
iA_j=\frac{1}{2}[(\Phi -1)X^*_i +c.c.]+\hat{\phi}_jJ\cdot \hat{r}B_1.
\end{equation}
Note that in the last term on the right, the unit vector $\hat{r}$ has been replaced by the unit vector $\hat{\phi}$.  Furthermore, the functions $\Phi_i$ now depend not just on $r$, but on the cylindrical coordinates $\rho ,\;z,\; \phi$, and the function $B_1$ is taken to be
\begin{equation}    %22%
B_1=\frac{1}{\rho}-MK_1(M\rho ),
\end{equation}
that is, the vortex factor which multiplies $\hat{\phi}$ in the center vortex taken by itself (equation (6)).  Of course, this choice for $B_1$ has cancellation of Dirac strings in its two parts, just as for the usual vortex.
There is a difference with the vortex in equation (6), however; in (21,22) the $SU(2)$ matrices enter as $J\cdot \hat{r}$, instead of $J_3$ as in (6).  This has the effect, as one readily checks, of giving Dirac-string singular field strengths proportional to $J_3$ but with {\em opposite} sign on either side of the nexus, as schematized in Fig. 1.  In the vortex, of course, the field strengths have the same sign of $J_3$ wherever one goes on the vortex.  Although this changes nothing about the asymptotic Wilson loop, which is indifferent to the sign of $J_3$, it is a gauge-invariant distinction.  There is no {\em regular} gauge transformation which changes $J_3$ to $J\cdot \hat{r}$; the gauge transformation which does the job, in equation (18), is certainly singular.  

The next step is to choose simple variational forms for the functions $\Phi_i$, which satisfy the appropriate boundary conditions at large distances from the nexus and vortex as expressed in equation (20), with $\beta =\pi$, and lead to vanishing gauge potential at the origin.  We will do this in analogy to such forms for the QCD sphaleron \cite{ct}.  The necessary $\phi$ dependence leads to:
\begin{eqnarray}   %23% 
\Phi_1 & = & F(\rho ,z)+G(\rho ,z)\cos \phi;\\ \nonumber
\Phi_2 & = & G(\rho ,z)\sin \phi.
\end{eqnarray}
The boundary conditions mentioned above lead to:
\begin{eqnarray}    %24%
\rho ,z\rightarrow \infty : & F\rightarrow 0,\;G\rightarrow -1 \\ \nonumber
\rho ,z\rightarrow 0: & F\rightarrow +1,\;G\rightarrow 0.
\end{eqnarray}
Simple functions obeying these boundary conditions are:
\begin{equation}   %25%
F=\frac{\lambda^2}{\lambda^2+r^2},\;G=\frac{-\rho r}{\lambda^2+\rho r}.
\end{equation}
Here $\lambda$ is a variational parameter.\footnote{For the QCD sphaleron, the choice $B_1$=0, $\Phi=(\lambda^2-r^2)(\lambda^2+r^2)$ gives a sphaleron mass only
1/2\% larger than the true mass.}  The appearance of $\rho$ in $G$ is required to prevent singularities from derivatives of $\phi$.

It only remains to calculate the energy or action from equation (1), and to drop the vortex terms.  The total energy has three terms:  The first term depends only on $\Phi_{1,2}$ (equation (23)); the second term depends only on the vortex wave function $B_1$ (equation (22)), and the third term is a cross term depending on all three functions.  We drop the vortex term, and the resultant energy is the nexus energy $E_n$. The $G^2$ part in equation (1) of the first term of the energy scales like $1/\lambda$, but the third (cross) term depends in a complicated way on both $M$ and $\lambda$ if we use literally the expression (22) for $B_1$, where the mass scale is $M$.  However, it turns out to make little difference if we replace $M$ in (22) by the variational parameter $1/\lambda$, in which case the $G^2$ part of the cross term also scales like $1/\lambda$.  For simplicity, we make this replacement.  The nexus energy $E_n$ is defined as the total energy less the vortex energy, and after inserting equations (21,22,23,25) in the energy functional and doing the integrals we find:  
\begin{equation}    %26%
E_n= 1.911\frac{\pi^2}{\lambda g^2}+2.197\frac{\pi^2M^2\lambda}{g^2};
\end{equation}
the $M^2\lambda $ term comes from the mass term of the energy.
Equation (26) has a minimum at $\lambda=0.933/M$, of value 3.22$(4\pi M/g^2)$.
This energy for the nexus is rather smaller than the conventional QCD sphaleron \cite{c76}, which is about 5.44$(4\pi M/g^2)$, and larger than that of the quasi-Abelian nexus, which is (see equation (14)) 2$\pi M/g^2$.

\begin{center}
{\bf B.  SU(3) non-Abelian nexus}
\end{center}

We wish to find for $SU(N)$ with $N>$2 the kind of nearly spherically-symmetric nexuses discussed above for $SU(2)$.  In particular, we desire to construct a nexus with $N$ center vortices, each of unit flux, meeting at the nexus.  It is clear from 
the properties of the $Q_i$ matrices (equation (4)) that this requires an embedding of $SU(2)$ other than the standard ones, such as $\lambda_1/2,\lambda_2/2,\lambda_3/2$ in $SU(3)$; these standard embeddings cannot span the entire Cartan subalgebra, as the set of $Q_i$ does.
Aside from the standard embedding of $SU(2)$ as a subalgebra of $SU(N)$, there is always another embedding, called the principal embedding \cite{k59}\footnote{I thank S. Ferrara for furnishing this reference.} which does span the entire Lie algebra of $SU(N)$.  The kinematics of this sort of embedding have already been used in Ref. \cite{cofn} for $SU(3)$.  Here we briefly review the principal embedding and the formation of a spherically-symmetric Cartan subalgebra, but stop short of a full-scale calculation of the properties of the resultant nexus.  The reason for not carrying out the full calculation is that the rotation generators we must use for $SU(3)$ are in the $J=1$ representation of $SU(2)$, and their trace is four times as large as the $J=1/2$ generators used in the standard embedding of a unitary group in the rotation group $SU(2)$.  As a result, the nexus energy is about four times as large as the conventional QCD sphaleron energy, and is so large as to appear to be unimportant.  Similarly, the CS number is four times as large, and has the value 2.  

In the principal embedding, the Lie algebra of
the fundamental $N\times N$ representation of $SU(N)$ can be formed from representations of $SU(2)$ with spin $J=(N-1)/2$ as follows.  Let $J_a$ be the generators of $SU(2)$ with this spin, and form further elements of the Lie algebra of
$SU(N)$ via:
\begin{equation}   %27%
M_{ab}J_aJ_b,\;M_{abc}J_aJ_bJ_c,\cdots
\end{equation}
where the $M$-tensors form a complete set of (numerical) symmetric and traceless tensors and the series terminates
at an $M$ with $N-1$ indices.  Clearly the set (27) plus the $J_a$ form a Lie algebra with $N^2-1$ elements.  One forms a spherically-symmetric version of the Cartan subalgebra by choosing a subset of the tensors $M_{abc\dots}$ as the appropriate symmetric and traceless combinations of the unit vector $\hat{r}_a$, {\it e.g.},
\begin{equation}     %28%
M_{ab}=\hat{r}_a\hat{r}_b-\frac{1}{3}\delta_{ab}.
\end{equation}
A spherically-symmetric Cartan subalgebra then is $J\cdot \hat{r},\;M_{ab}\hat{r}_a\hat{r}_b,\dots$.

Now we specialize to $SU(3)$.  For this group the principal embedding is the familiar ``nuclear physics" decomposition into $O(3)$ plus a quadrupole tensor.  In this group one has $\{J_1,J_2,J_3\}=\{\lambda_7,-\lambda_5,\lambda_2\}$ in terms of the Gell-Mann matrices and we normalize the symmetric and traceless quadrupole generators as:
\begin{equation}     %29%
T_{ab}\equiv \frac{1}{2}\{J_a,J_b\} -\frac{2}{3}I\delta_{ab};\;\sum_aT_{aa}=0.
\end{equation}
The diagonal elements span the Cartan subalgebra.  Explicitly:
\begin{eqnarray}    %30% 
T_{11}=diag(-2/3,1/3,1/3);\;T_{22}=diag(1/3,-2/3,1/3);\nonumber \\
T_{33}=diag(1/3,1/3,-2/3).
\end{eqnarray}
The Lie algebra takes the form:
\begin{equation}     %31%
[J_a,T_{bc}]=i\epsilon_{abd}T_{dc}+i\epsilon_{acd}T_{bd};
\end{equation}
\begin{equation}     %32%
[T_{ab},T_{cd}]=\frac{i}{4}J_e(\epsilon_{ace}\delta_{bd}+
\epsilon_{bce}\delta_{ad}+\epsilon_{ade}\delta_{bc}+
\epsilon_{bde}\delta_{ac}).
\end{equation}
The $T_{ab}$ are normalized according to:
\begin{equation}       %33%
TrT_{ab}T_{cd}=\frac{1}{2}(\delta_{ac}\delta_{bd}+\delta_{bc}\delta_{ad}
-\frac{2}{3}\delta_{ab}\delta_{cd}).
\end{equation}

The spherical basis for the Cartan subalgebra is $J\cdot \hat{r}$ and
\begin{equation}     %34%
Q\equiv \hat{r}_a\hat{r}_bT_{ab}.
\end{equation}
The most general spherically-symmetric form of the gauge potential $A_i$ for 
$SU(3)$ can be read off from the kinematic structure of the pure-gauge potential
\begin{equation}    %35%
A_i(\vec{r})=V\partial_iV^{-1},\;V=\exp i(\alpha (r)Q +\beta (r)J\cdot \hat{r}).
\end{equation}
This yields:
\begin{equation}     %36% 
iA_j=B_1\hat{r}_jQ+C_1\hat{r}_jJ\cdot \hat{r}
+\frac{1}{2}[(\Phi -1)X^*_j+c.c.]
\end{equation}
where $\Phi$ is a {\em complex doublet} scalar and $X_j$ is the
kinematic complex doublet
\begin{equation}     %37%
 \left( \begin{array}{l}
J_j-\hat{r}_jJ\cdot \hat{r}+i\epsilon_{jkl}J_k\hat{r}_l\\
K_j+i\epsilon_{jkl}K_k\hat{r}_l
\end{array} \right) 
\end{equation}
and
\begin{equation}     %38%
K_i=r\partial_iQ=2(T_{ia}\hat{r}_a-\hat{r}_iQ).
\end{equation}
Here the functions $B_1,C_1,\Phi$ are all functions of $r$.
This is, of course, the same basis as used by Corrigan {\it et al.} \cite{cofn}, but in a different notation.
It can be extended to time-dependent functions by adding $A_4$ as a linear combination 
\begin{equation}      %39%
iA_4=B_2Q+C_2J\cdot \hat{r}.
\end{equation}  

Under a gauge transformation given by $V$ of (35), one finds that 
\begin{equation}    %40%
\phi \rightarrow e^{-i(\beta +\sigma_1\alpha)}\phi
\end{equation}
while $B_{\gamma},C_{\gamma}(\gamma=1,2)$ are changed by gradients of $\alpha ,\beta$ as Abelian gauge fields would be.  So the interpretation, similar to that of the corresponding Witten spherical ansatz for $SU(2)$, is that of a complex d=2 scalar field with two $U(1)$ gauge symmetries as indicated by (40), with $B,C$ the corresponding gauge potentials.

Next we deform the spherically-symmetric ansatz as we did for $SU(2)$, by constructing an $SU(3)$ equivalent of equation (19).  This requires first a specification of some angular functions whose gradient gives the long-range pure-gauge part of the vortices (the $\Delta_0$ term in equation (2)).  The necessary generalization uses what is known, in the older electromagnetic literature, as the magnetic potential $\Phi$.  Corresponding to any closed curve $\Gamma$, the magnetic potential is defined as:
\begin{equation}   %41%
\Phi_{\Gamma}=\frac{1}{2}\epsilon_{ijk}\int_S d\sigma_{ij}\partial_k\frac{1}{|\vec{x}-\vec{z}|}
\end{equation}  
where the surface $S$ has the curve $\Gamma$ as its boundary.  The gradient of $\Phi$ gives the $\Delta_0$ term in (2), with a $2\pi$ jump giving the Dirac-string singular magnetic field whenever the surface $S$ is pierced by a closed loop linked to $\Gamma$.  

We choose three curves, much as shown in Fig. 6(b), which has, as it must, two $SU(3)$ nexuses joined by three thick vortices.   There is also a line along which all the three strings 1,2,and 3 coincide, allowing for closure of these three curves.  To construct a single $SU(3)$ nexus in isolation, imagine that one of the two nexuses in the figure is taken to infinity.  

At large distances from the nexus and its attached vortices, the gauge potential must approach a pure gauge ({\it cf.} equation (20) for $SU(2)$).  We choose this gauge to be:
\begin{equation}     %42%
A_i\rightarrow R\partial_iR^{-1},
\end{equation}
\begin{equation}    %43%
R=\exp i[Q(\Phi_3-\frac{1}{2}\Phi_1-\frac{1}{2}\Phi_2)+\frac{1}{2}J\cdot \hat{r}(\Phi_2 -\Phi_1)].
\end{equation}
One easily sees that along the line where the three strings coincide there is no $2\pi$ jump in the exponent of $R$, and no Dirac string.  On the other hand, the jump associated with any string taken by itself leads to a unit-flux Dirac string.  For example, the $2\pi$ jump around string 1 leads to a flux factor expressed in
\begin{equation}    %44%
R\rightarrow \exp \{ -i\pi [Q+J\cdot \hat{r}]\} = \exp \frac{2\pi i}{3}.
\end{equation}
If the strings are chosen to lie along the positive coordinate axes, as in the quasi-Abelian $SU(3)$ vortex discussed above, the matrix $R$ is unitarily equivalent to the same matrix with the coefficients of the $\Phi_a$ replaced by $T_{aa}$; this is the analog of the construction of the gauge matrix $W$ for $SU(2)$ in equation (19).

\begin{center}
{\bf V.  SUMMARY AND CONCLUSIONS}
\end{center}

In $SU(N)$ we have partially (fully, for the quasi-Abelian case) constructed some new solitons for the center-vortex picture, solitons which have up to $N$ thick vortices extending from them, with total magnetic flux of zero (mod $N$).  These solitons can be thought of as composed of $N$ monopoles of unit flux.  Such monopoles are unacceptable in isolation, as they would come with singular naked Dirac strings, but in the nexus these Dirac strings cancel.  For $N\geq 3$ these solitons, if entropically favored, would lead to a vortex vacuum analogous to a branched-polymer gel.  There is, however, no effect from this branching on the fundamental properties of the area law for Wilson loops.  Some of the nexuses carry CS number (or topological charge in four dimensions).  In fact, vortices themselves carry CS number or topological charge, quantized in units of $1/N$.  Such quantization is quite natural from the point of view of solving the $U(1)$ problem\cite{w79}, but is not natural from the viewpoint of periodicity of the $\theta$ angle in units of 2$\pi$.  However, these requirements can be reconciled, and we will discuss this in the center-vortex picture in a later publication.  

Even though the nexuses may not play a large dynamical role in QCD (at least at $\theta=0$), it is worthwhile looking for them on the lattice, supplied with appropriate boundary conditions, as a test of the underlying properties of center vortices.  It might be worthwhile to list here some of the other properties of the center vortex picture which are accessible to lattice calculations and other tests.  First, the baryonic Wilson-loop area law is predicted\cite{c96} to be a $\Delta$-law, not a Y-law. Second, the fundamental premise of the center-vortex picture is that confinement comes from a non-trivial group center.  There are certain exceptional groups (the simplest is $G_2$) which have only a trivial center, and it would be interesting to simulate them on the lattice to see if they confine.  And third, there are tests yet to be devised which probe the interaction of vortices and the $\theta$ angle.

\begin{center}
{\bf ACKNOWLEDGMENTS}
\end{center}

This work was supported in part by the National Science Foundation under grant PHY9531023.

\newpage
\begin{center}
{\bf FIGURE CAPTIONS}
\end{center}

\bigskip

Fig. 1.  An isolated $SU(2)$ nexus (circle) and unit-flux vortices (lines), with field strengths reversing at the nexus.

Fig. 2.  An isolated $SU(3)$ nexus-vortex combination. 

Fig. 3.  $SU(2)$ nexuses in a closed vortex.

Fig. 4.  Possible configurations for $SU(3)$ nexuses and vortices.

Fig. 5.  An isolated $SU(N)$ vortex-nexus combination, with up to $N$ vortices meeting at the nexus; the sum of fluxes must vanish (mod $N$).

Fig. 6.  Depiction of $SU(3)$ vortex strings as closed Dirac strings.  The double string in (a) is physical, but not in (b), since it carries no flux or energy.  In (a) the double string is unstable to separation for the Abelian nexus.

\newpage
\epsfig{file=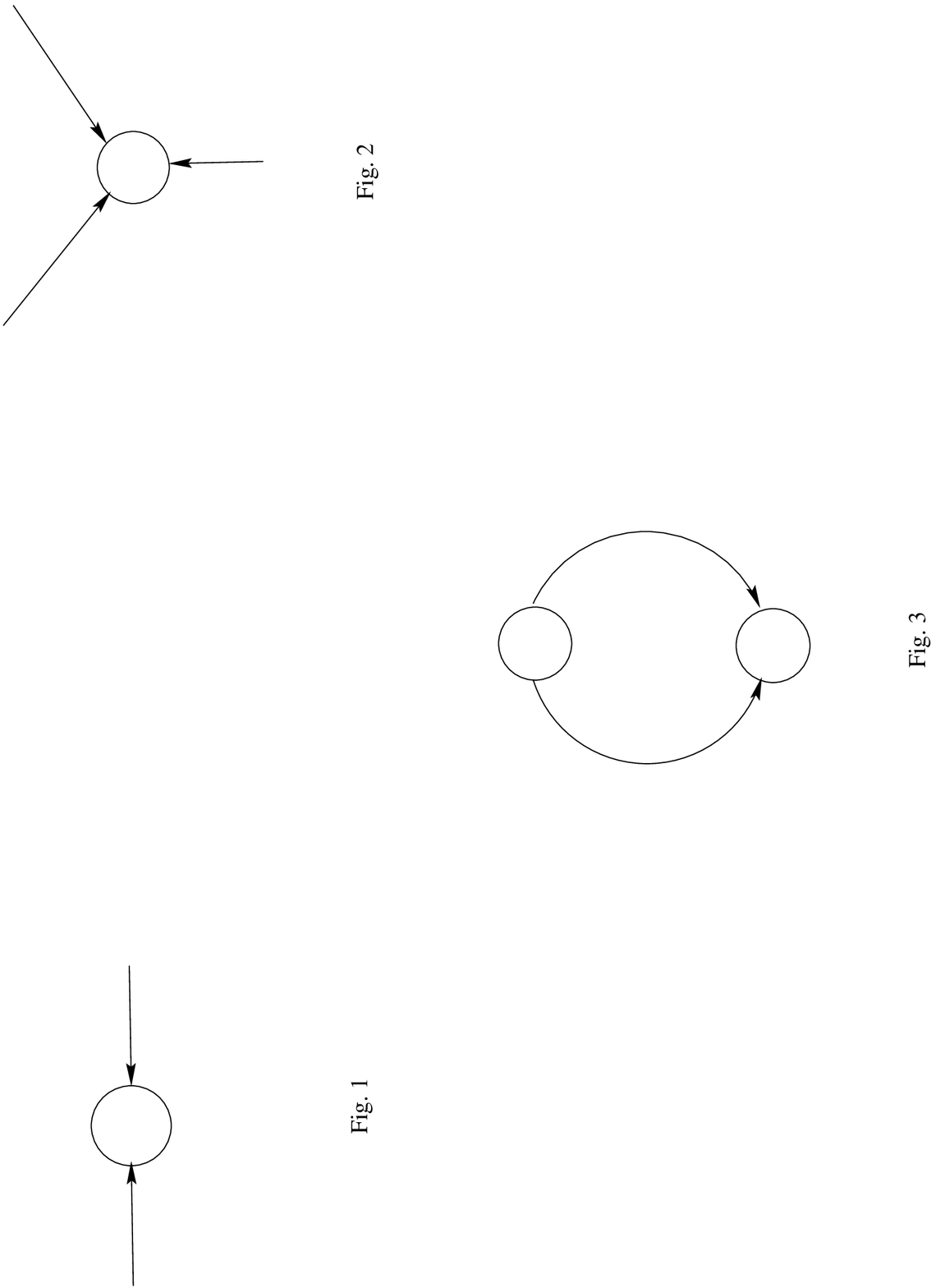,clip=}
\newpage
\epsfig{file=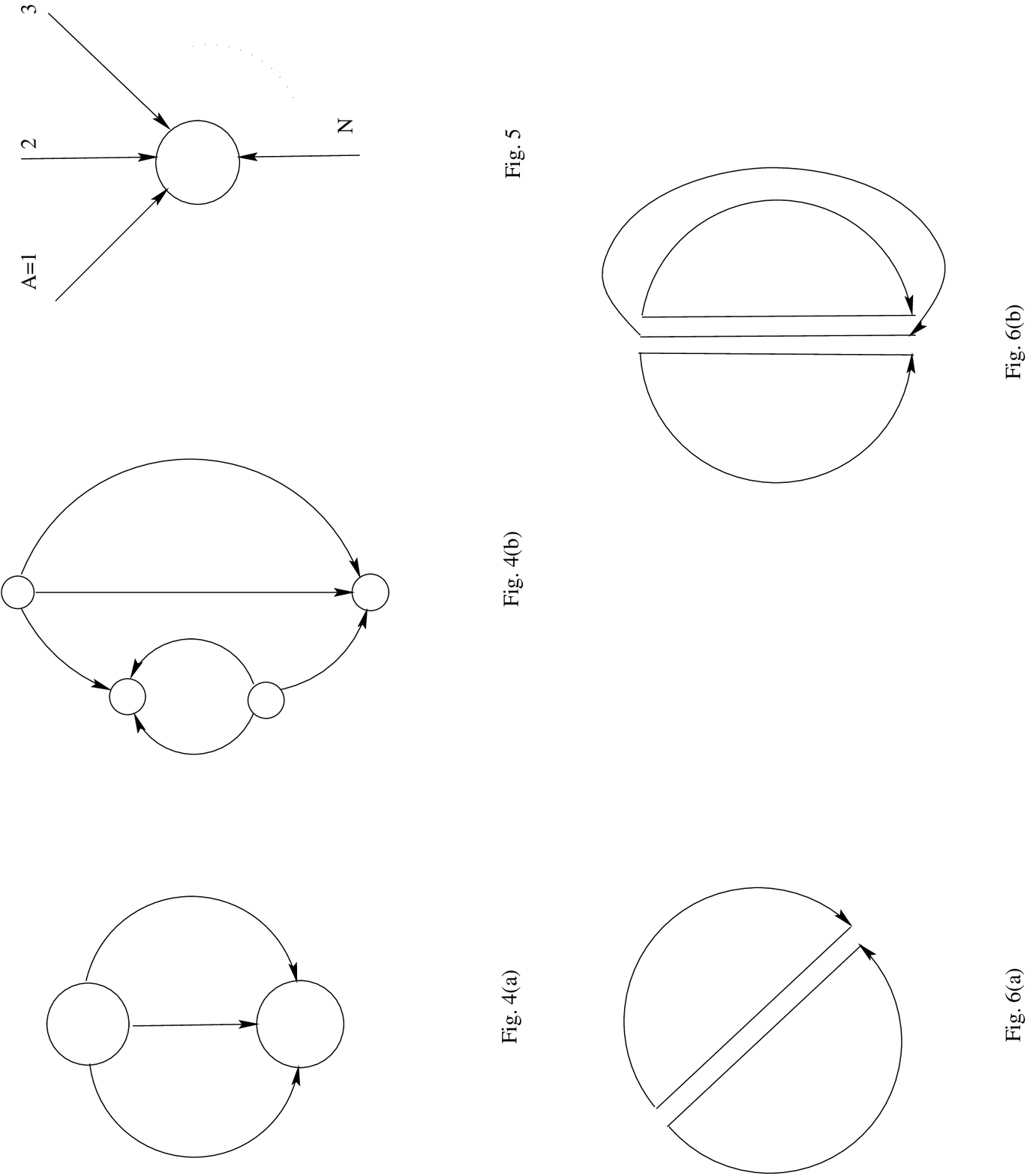,clip=}


\newpage
\begin{thebibliography}{99}
\bibitem{c79} J. M. Cornwall, Nucl. Phys. B{\bf 157}, 392 (1979).
\bibitem{th79} G. 't Hooft, Nucl. Phys. B{\bf 153}, 141 (1979).
\bibitem{mp} G. Mack and V. B. Petkova, Ann. Phys. (NY) {\bf 132}, 442 (1979); {\it ibid.} {\bf 125}, 117 (1980); Z. Phys. C {\bf 12}, 177 (1982).
\bibitem{no79} H. B. Nielsen and P. Olesen, Nucl. Phys. B {\bf 160}, 380 (1979); J. Ambj\o rn and P. Olesen, Nucl. Phys. B {\bf 170}, 60 (1980); {\it ibid.} 225.
\bibitem{y80} L. Yaffe, Phys. Rev. D{\bf 221}, 1574 (1980).
\bibitem{t93} E. T. Tomboulis, Phys. Lett. B {\bf 303}, 103 (1993); Nucl. Phys. B (Proc. Suppl.) {\bf 34}, 192 (1994).
\bibitem{kt} T. G. Kov\'acs and E. T. Tomboulis, Nucl. Phys. B (Proc. Suppl.) {\bf 53}, 509 (1997); Nucl. Phys. B (Proc. Suppl.) {\bf 63}, 534 (1998); Phys Rev. D {\bf 57}, 4054 (1998).
\bibitem{fgo}  L. Del Debbio, M. Faber, J. Greensite, and \u{S}. Olejn\'{\i}k, Phys. Rev. D {\bf 55}, 2298 (1997); presentation at the 31st International Symposium on the Theory of Elementary Particles, Buckow,and 7th Workshop on Lattice Field Theory, Vienna (preprint hep-lat/9802003).
\bibitem{getal} L. Del Debbio, M. Faber, J. Greensite, and \u{S}. Olejn\'{\i}k, presentation at the Nato Workshop ``New Developments in Quantum Field Theory", Zakopane, Poland, June 1997 (hep-lat/9708023).
\bibitem{kl} K. Langfeld, H. Reinhardt, and O. Tennert, Phys. Lett. B{\bf 419}, 317 (1998).
\bibitem{b82} C. Bernard, Phys. Lett. B{\bf 108}, 431 (1982).
\bibitem{pt} G. I. Poulis and H. D. Trottier, Phys. Lett. B{\bf 400}, 358 (1997).
\bibitem{c83} J. M. Cornwall, in Progress in Physics, V. 8, {\it Workshop on Non-Perturbative Quantum Chromodynamics}, Proceedings of the Conference, Stillwater, Oklahoma, 1983, edited by K.A. Milton and M. A. Samuel (Birkh\"auser, Boston, 1983).
\bibitem{c98} J. M. Cornwall, preprint UCLA/97/TEP/30 (hep-th/9712248, unpublished).
\bibitem{gad} M. Faber, J. Greensite, and \u{S}. Olejn\'{\i}k, preprint
hep-lat/9710039 (October 1997, unpublished).
\bibitem{cofn} E. Corrigan, D. I. Olive, O. B. Fairlie, and J. Nuyts, Nucl. Phys. B{\bf 106}, 475 (1976).
\bibitem{c76} J. M. Cornwall, in {\it Deeper Pathways in High-Energy Physics}, Proceeding of the Conference, Coral Gables, Florida, 1977, edited by B. Kursonoglu {\it et al.} (Plenum, New York, 1977), p. 683.
\bibitem{ct} J. M. Cornwall and G. Tiktopoulos, Phys. Lett. B {\bf 181}, 353 (1986).  
\bibitem{c94} J. M. Cornwall, in {\it Unified Symmetry in the Small and in the Large}, Proceedings of the Conference, Coral Gables, Florida, 1994, edited by B. Kursonoglu {\it et al.} (Plenum, New York, 1995), p. 243. 
\bibitem{c82} J. M. Cornwall, Phys. Rev. D{\bf 26}, 1453 (1982).
\bibitem{la} M. Lavelle, Phys. Rev. D{\bf 44}, R26 (1991).
\bibitem{k59} B. Kostant, Am. J. Math. {\bf 81}, 973 (1959).
\bibitem{w79} E. Witten, Nucl. Phys. B{\bf 149}, 285 (1979).
\bibitem{c96} J. M. Cornwall, Phys. Rev. D {\bf 54}, 6527 (1996).
\end{thebibliography}
\end{document}